\input{aipcheck}

\documentclass[
    ,final            
  ]
  {aipproc}

\layoutstyle{6x9}

\begin{document}

\title{Toward the Ab-initio Description of Medium Mass Nuclei}


\author{C. Barbieri}{
  address={Department of Physics, University of Surrey, Guildford GU2 7XH, UK}
}

\author{A. Cipollone}{
  address={Department of Physics, University of Surrey, Guildford GU2 7XH, UK},
  altaddress={Dipartimento di Fisica, Universit\`a "Sapienza'', I-00185 Roma, Italy},
  altaddress={INFN, Sezione di Roma, Piazzale Aldo Moro 2, I-00185 Roma, Italy}
}

\author{V. Som\`a}{
  address={Institut f\"ur Kernphysik, Technische Universit\"at Darmstadt, 64289 Darmstadt, Germany},
  altaddress={ExtreMe Matter Institute EMMI, GSI Helmholtzzentrum f\"ur Schwerionenforschung GmbH, 64291 Darmstadt, Germany}
}

\author{T. Duguet}{
  address={CEA-Saclay, IRFU/Service de Physique Nucl\'eaire, F-91191 Gif-sur-Yvette, France }, 
  altaddress={National Superconducting Cyclotron Laboratory and Department of Physics and Astronomy,
Michigan State University, East Lansing, MI 48824, USA}
}

\author{P. Navr\'atil}{
  address={TRIUMF, 4004 Wesbrook Mall, Vancouver, British Columbia, V6T 2A3, Canada }
}

\begin{abstract}
 As ab-initio calculations of atomic nuclei enter the A=40-100 mass range, a great challenge is how 
to approach the vast majority of open-shell (degenerate) isotopes. 
 We add realistic three-nucleon interactions to the state of the art many-body Green's function
theory of closed-shells, and find that physics of neutron driplines is reproduced with very good
quality. Further, we introduce the Gorkov formalism to extend {\em ab-initio} theory to semi-magic,
fully open-shell, isotopes. Proof-of-principle calculations for $^{44}$Ca and $^{74}$Ni confirm
that this approach is indeed feasible. 
Combining these two advances (open-shells and three-nucleon interactions) requires longer,
technical, work but it is otherwise within reach.

\end{abstract}

\maketitle


\paragraph{Introduction}
Microscopic first principle predictions of atomic nuclei are highly desirable 
since they can unambiguously guide research of exotic isotopes.
These could also help in constraining extrapolations
to higher mass regions~\cite{Erle.2012} and to extreme proton-neutron asymmetries~\cite{Wald.2011},
including regions close to the driplines where experimental data is unlikely
to become available in the foreseeable future.

Ab-initio methods such as coupled-cluster (CC)~\cite{Jans.2011}, in-medium similarity renormalization group (IMSRG)~\cite{Tsuk.2011} or Dyson self-consistent Green's function~\cite{DiB.04,FTDA} (Dyson-SCGF) have accessed medium-mass nuclei up to A$\sim$60 on the basis of realistic two-nucleon (2N) interactions.
However, it has become clear that three-nucleon forces (3NFs) play a major role in determining crucial features of exotic isotopes, such as the evolution of magic numbers and the position of driplines~\cite{Otsu.2010,Holt.2012,Hage.2012}. Realistic 2N and 3N interactions based on chiral perturbation theory have recently been evolved to low cutoffs, retaining both induced and pre-existing 3NFs~\cite{Jurg.2009,Roth.2012}. Proper implementations of similar hamiltonians within the above many-body theories will be required to eventually achieve quantitative predictions of medium-mass isotopes.

A second (and major) challenge to ab-initio theory is that current implementations of the above methods are  limited to doubly closed (sub-)shell nuclei and their immediate neighbors~\cite{Jans.2011,FTDA}. As one increases the nuclear mass, longer chains of truly open-shell nuclei connecting isolated doubly closed-shell ones emerge and cannot be accessed with existing approaches. Many-body techniques that could tackle genuine (at least) singly open-shell systems would immediately extend the reach of ab-initio studies from a few tens to several hundreds of mid-mass nuclei. A manageable way to fill this gap was recently proposed in Refs.~\cite{papI,G.PRL} by extending SCGF to Gorkov formalism and will be discussed in the following. 
This talk reports on recent progress on the above topics.

\paragraph{Three-nucleon interactions}

\begin{figure}
\includegraphics[height=.19\textheight]{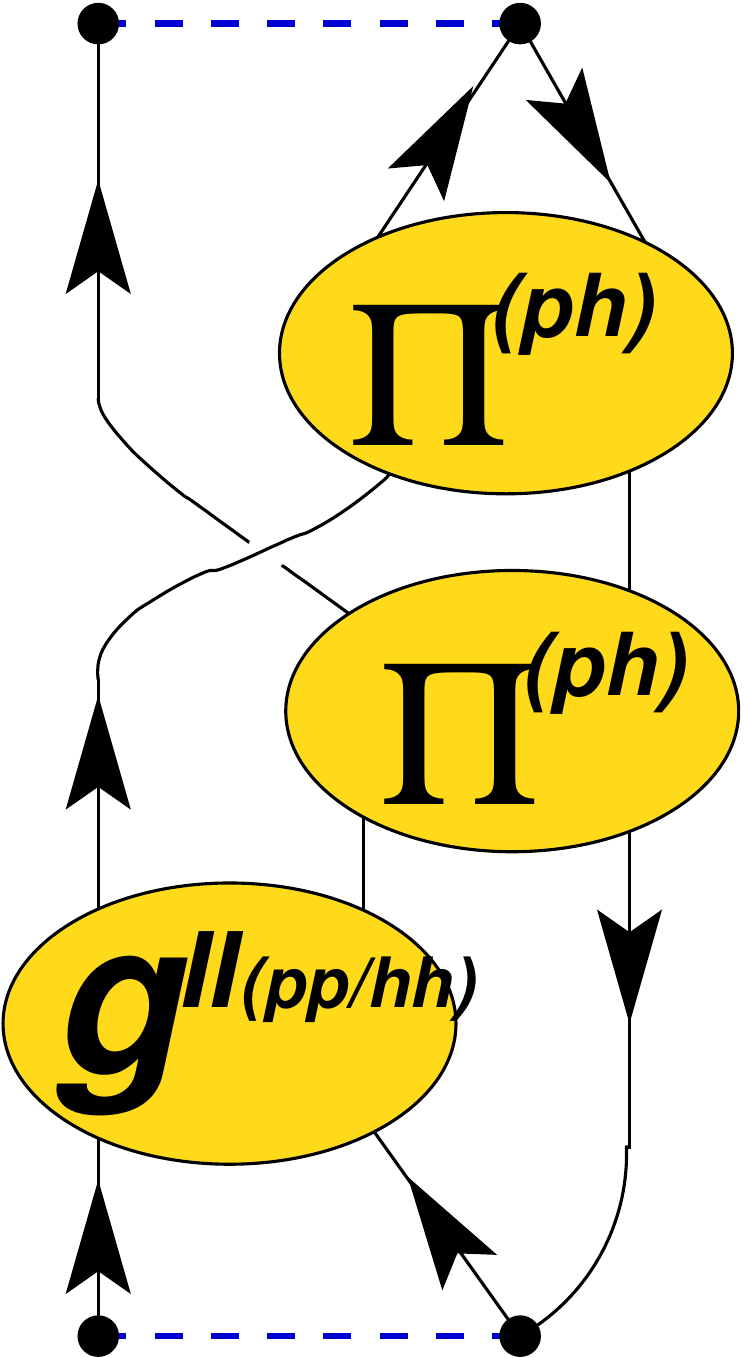}
\hspace{0.5in}
\includegraphics[height=.23\textheight]{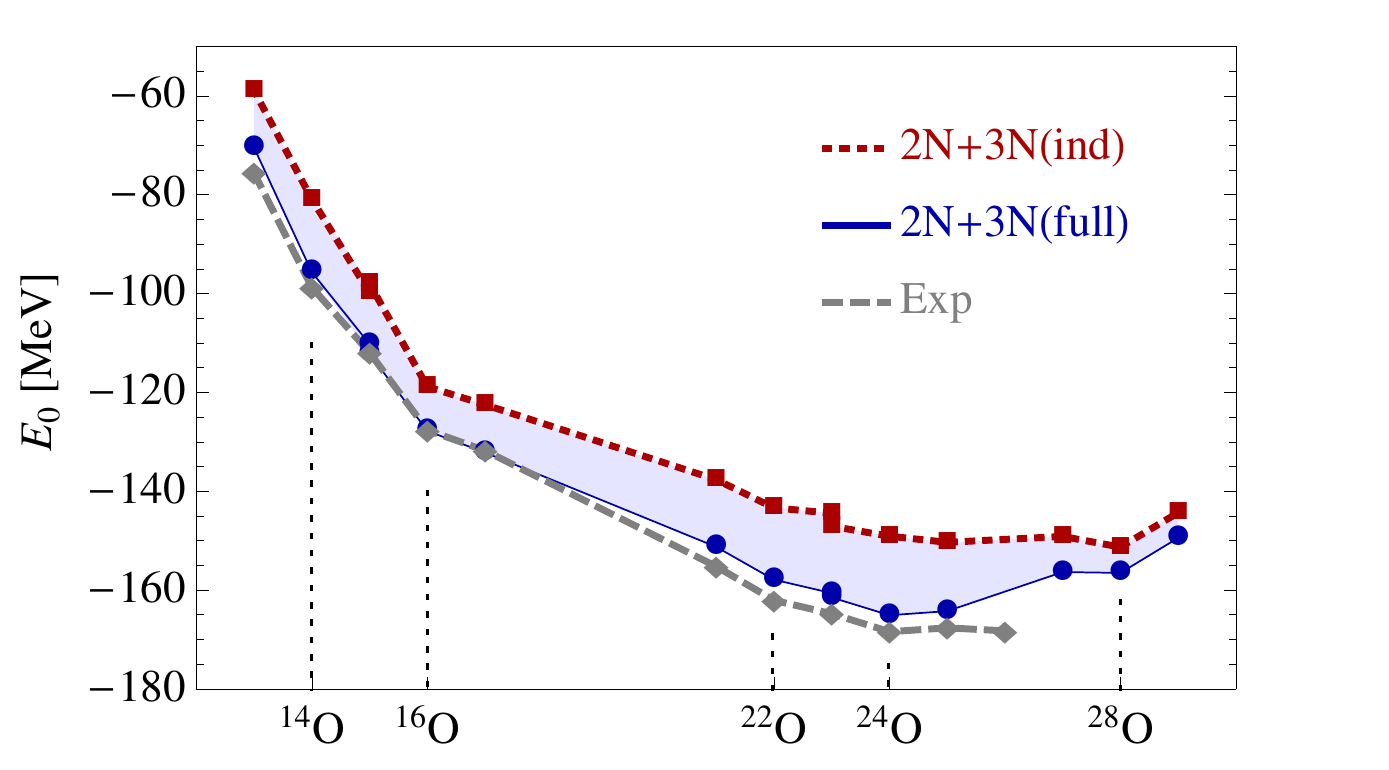}
  \caption{{\em Left:} Example of a particle-vibration coupling diagram  entering the FTDA/ADC(3) self energy.   {\em Right:} Binding energies of oxygen isotopes obtained from a SRG evolved NN+3N interactions with cutoff $\lambda$=1.88 fm$^{-1}$.  Squares (dots) refer to
induced-only (full) three-nucleon interactions and are compared to experiment (diamonds). Binding energies of even-N isotopes are obtained through the
corrected Koltum sum rule (\ref{eq:koltun}). Odd-N energies are inferred from addition and separation energies, as obtained from the poles of the propagator~(\ref{eq:g1}).
    \label{fig:FTDA} }
\end{figure}
 
We employ Green's function (or propagator) theory, where the object of interest is the single particle propagator~\cite{FetWal},
\begin{equation}
 g_{\alpha \beta}(\omega) ~=~ 
 \sum_n  \frac{ 
          \langle {\Psi^A_0}     \vert c_\alpha        \vert {\Psi^{A+1}_n} \rangle
          \langle {\Psi^{A+1}_n} \vert c^{\dag}_\beta  \vert {\Psi^A_0} \rangle
              }{\omega - (E^{A+1}_n - E^A_0) + i \eta }  ~+~
 \sum_k \frac{
          \langle {\Psi^A_0}     \vert c^{\dag}_\beta  \vert {\Psi^{A-1}_k} \rangle
          \langle {\Psi^{A-1}_k} \vert c_\alpha        \vert {\Psi^A_0} \rangle
             }{\omega - (E^A_0 - E^{A-1}_k) - i \eta } \; ,
\label{eq:g1}
\end{equation}
where $\vert\Psi^{A+1}_n\rangle$, $\vert\Psi^{A-1}_k\rangle$ 
are the eigenstates, and $E^{A+1}_n$, $E^{A-1}_k$ the eigenenergies of the 
($A\pm1$)-nucleon system. Therefore, the poles of the propagator reflect  
nucleon addition and separation energies.  The propagator is calculated for
finite closed-shell nuclei by first solving spherical Hartree-Fock (HF) equations.
The HF state is then used as a reference state for the Faddeev Tamm-Dancoff (FTDA) 
method [a.k.a. ADC(3)] of Refs.~\cite{FTDA}. The FTDA method completely accounts
for particle-vibration diagrams as  shown in Fig.~\ref{fig:FTDA}.

We employ the intrinsic hamiltonian $H_{int}\equiv H-T_{c.m.}=\hat{U}+\hat{V}+\hat{W}$, where the 
kinetic energy of the center of mass has been subtracted and $\hat U$, $\hat V$ and $\hat{W}$ are the
one-, two-, and three-nucleon components, respectively.  Form this, we generate  one- and 
two-nucleon density dependent interactions with matrix elements,
\begin{eqnarray}
u^{(3NF)}_{\alpha \beta}  & =& \frac{1}{2}
 \sum_{\gamma \, \sigma \, \mu \, \nu}\frac {1}{(2 \pi i)^2} \int_{C \uparrow} d \omega_1 \; \int_{C \uparrow} d \omega_2 \;\;
     w_{\alpha \mu \nu, \beta \gamma \sigma } \, g_{\gamma \mu}(\omega_1)  \, g_{\sigma \nu}(\omega_2)  \; ,
\nonumber \\
v^{(3NF)}_{\alpha \beta, \gamma \delta} & =& \sum_{\mu \, \nu}\frac{1}{2 \pi i} \int_{C \uparrow} d \omega \;\;
     w_{\alpha \beta \mu, \gamma \delta \nu} \, g_{\nu \mu}(\omega)  \; .
\label{eq:effW}
\end{eqnarray}
These definition extend the normal ordering approach of Ref.~\cite{Roth.2012} by contracting with fully correlated
propagators, as opposed to a mean-field reference state. The matrix elements $u^{(3NF)}_{\alpha \beta}$
and $v^{(3NF)}_{\alpha \beta, \gamma \delta}$ are then added to the existing 1N and 2N forces with the 
caveat that only interaction irreducible diagrams are retained to ensure the correct symmetry factors
in the diagrammatic expansion~\cite{Carb.IP}.


After obtaining the sp propagator $g(\omega)$ the total binding energy can be calculated 
as usual through the Koltun sum rule which---due the the presence of 3NF---acquires the
corrected form
\begin{equation}
\label{eq:koltun}
E^A_0 = \sum_{\alpha \, \beta} \frac{1}{4 \pi i} \int_{C \uparrow} d \omega  \; \;
\left[ u_{\alpha \beta}  +  \omega \delta_{\alpha \beta} \right] \, g_{\beta \alpha}(\omega)   
~ - ~ \frac{1}{2}\langle {\Psi^A_0}     \vert \hat{W}     \vert {\Psi^A_0} \rangle  \; .
\end{equation}
Eq.~(\ref{eq:koltun}) is still an exact equation. However, it requires to evaluate the
expectation value of the 3NF part of the hamiltonian $< \hat{W}>$ which is calculated
here to first order in $\hat{W}$.

Calculations for closed sub-shell oxygen isotopes were performed for the chiral N$^3$LO 2NF~\cite{Mach.2003}
and N$^2$LO 3NF~\cite{Navr.2007} with the cutoff of 400 MeV as introduced in Ref.~\cite{Roth.2012}. These were 
evolved to a cutoff $\lambda=1.88$\,fm$^{-1}$ using free-space similarity renormalization
group (SRG)~\cite{Scott.PPNP}. We employed large model spaces of
up to 12 harmonic oscillator shells with frequency $\hbar\omega$=20\,MeV.
Results for the induced 3NF are obtained from the SRG evolution of the original 2NF only and are 
indicated by red squares in Fig.~\ref{fig:FTDA}. These are to be considered analogous to predictions of the
sole N$^3$LO 2NF and systematically under bind the oxygen isotopes. 
Adding full 3NFs, that include in particular the two-pion exchange Fujita-Miyazawa contribution, reproduces
experimental binding energies throughout the isotopic chain and the location of the neutron dripline.
Table~\ref{tab:radii} shows that although SRG evolved 2NFs alone underestimate the nuclear radii,
results improve with the inclusion of 3NFs.

\begin{table}[t]
\begin{tabular}{rcccccccc}
\hline
  & & \tablehead{1}{c}{b}{2NF only}
  & & \tablehead{1}{c}{b}{2+3NF(ind.)}
  & & \tablehead{1}{c}{b}{2+3NF(full)}
  & & \tablehead{1}{c}{b}{Experiment}   \\
\hline
$^{16}$O: & & 2.10  & & 2. 41 & & 2.38  & & 2.718$\pm$0.210~\cite{Wohl.1981} \\   
$^{44}$Ca: & & 2.48  & & 2.93  & & 2.94  & & 3.520$\pm$0.005~\cite{Schu.75} \\     
\hline
\end{tabular}
\caption{Predicted matter radii (in fm) for $^{16}$O and $^{44}$Ca form SRG evolved 2N-only interactions
and by including induced and full 3NF. Experiment are charge radii.}
\label{tab:radii}
\end{table}

%

\paragraph{Gorkov formalism for open-shell isotopes}

\begin{figure}[t]
  \hbox{
  \includegraphics[width=.445\textwidth]{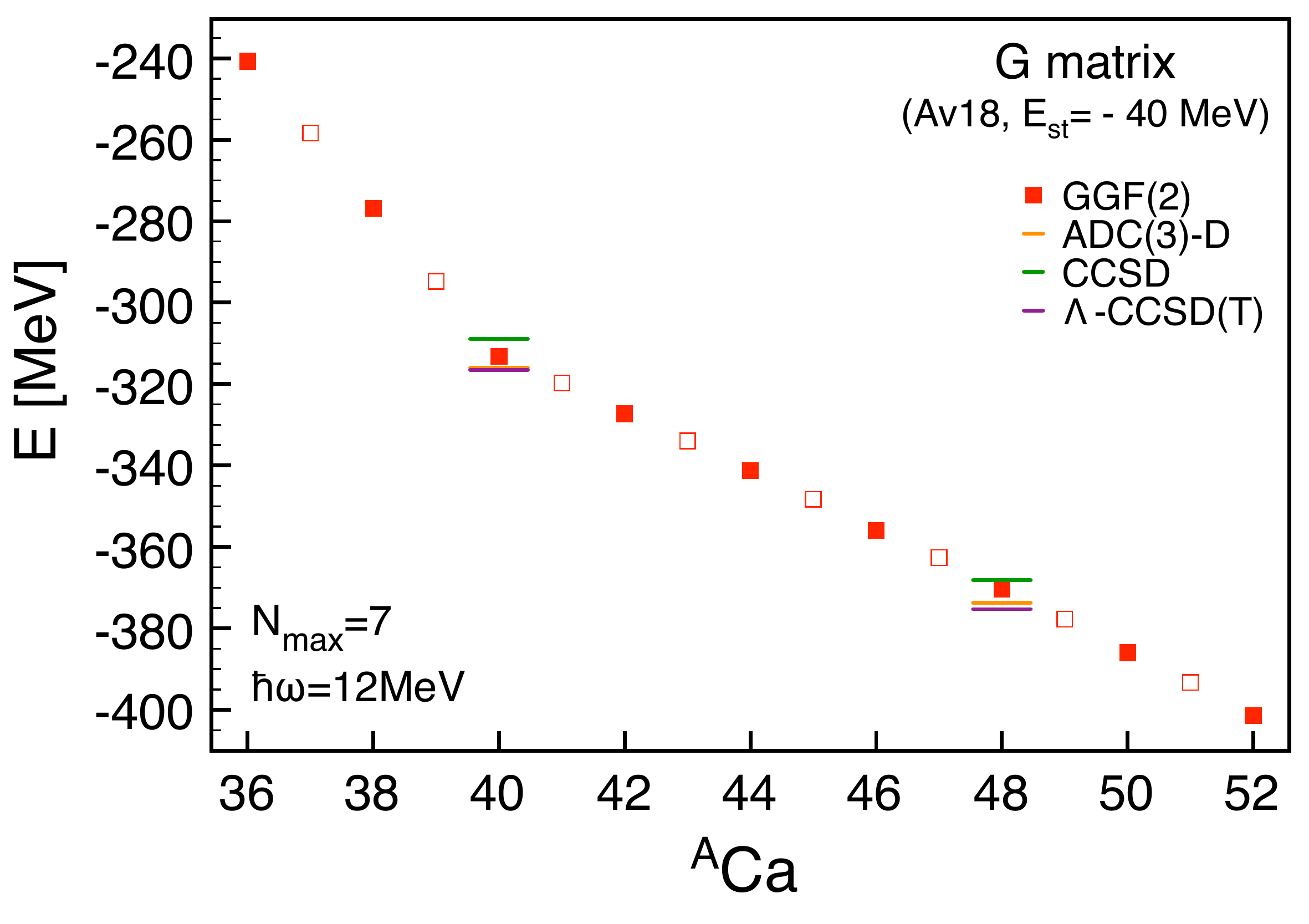}
  }
  \hbox{
  \includegraphics[width=.51\textwidth]{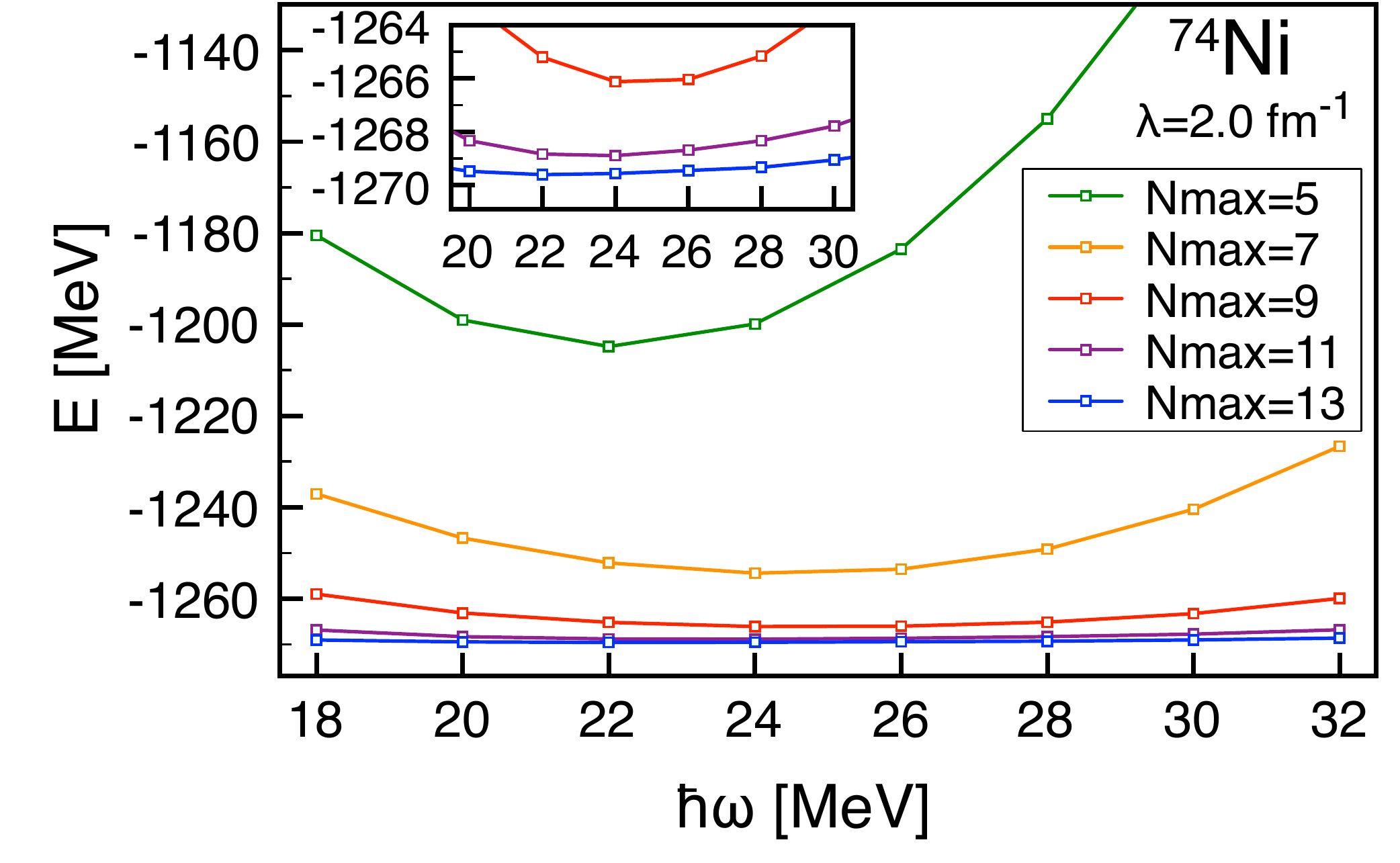}
  \spaceforfigure{.5cm}{1.cm}
  }
  \caption{  Results obtained from second-order Gorkov-SCGF {\em Left}: Binding energies of Ca isotopes for a fixed model space of eight shells.  Gorkov propagators are calculated for even-A (filled symbols) while odd-A results (open symbols) are computed according to Ref. \cite{Dugu.2002}.    Closed-shell $^{40}$Ca and $^{48}$Ca are compared to CC and Dyson-SCGF results.
{\em Right}:  Binding energy $^{74}$Ni as a function of the harmonic oscillator spacing $\hbar\omega$ and for an increasing size $N_{max}$ of the single-particle model space.The insert shows a zoom on the most converged results. 
 \label{Fig:Gorkov} }
\end{figure}

The Gorkov's approach handles intrinsic degeneracies of open shell systems
by allowing the breaking of particle number symmetry. One considers the grand
canonical hamiltonian $\Omega_{int} = H_{int} - \mu_p\hat{Z} - \mu_n\hat{N}$
and constrains expectation values of proton and neutron number operators to the
expected values.
This allows defining a
superfluid state which already accounts for pairing correlation and can be used
as reference for Green's function diagrammatic expansion. 
The formalism for Gorkov self-consistent Green's function (Gorkov-SCGF) theory up to 
second order in the self-energy has been worked out in full in Ref.~\cite{papI}, for 
2N interactions only. First results are reported in~\cite{G.PRL}.

The left panel in figure~\ref{Fig:Gorkov} displays the binding energies of calcium isotopes and  compares them to single-reference CC and Dyson-SCGF for closed-shell $^{40}$Ca and $^{48}$Ca, using a G-matrix interaction at a fixed starting energy. 
Already at second-order in the self-energy, Gorkov-SCGF can provide comparable accuracy to CC singles and doubles (CCSD). Higher order corrections introduced by triples [$\Lambda$-CCSD(T)] are closely reproduced by Dyson-SCGF in the FTDA/ADC(3) approximation. Since the extension of Gorkov's formalism to ADC(3) schemes is within computational reach, this gives confidence that Gorkov-SCGF  calculations can be improved to desired accuracy. The right panel displays good convergence properties, with respect to the model space, for isotopes as heavy as $^{74}$Ni, using 2N SRG interactions with cutoff $\lambda$=2.0\,fm$^{-1}$.   These findings demonstrate the feasibility of first-principle calculations along full isotopic chains based on Gorkov-SCGF theory.

We further consider $^{44}$Ca with the SRG, $\lambda$=2.0\,fm$^{-1}$, interaction and add a crude estimate of 3NF by calculating the normal self-energy in a filling approximation.  Full 3NFs are found to shift the neutron Fermi energy to -8.69 MeV, fairly close to the experiment. The neutron shell gap between the $0f_{7/2}$ and $0d_{3/2}$ is reduced from 12.9\,MeV (2NF only) to 7.2\,MeV (full 3NF). The gap between the centroids of their distributions~\cite{Dugu.2012} is 9.3\,MeV, in agreement with data driven predictions of Ref.~\cite{Char.2007}. The calculated  r.m.s. matter radius of 2.94~fm (Tab.~\ref{tab:radii}) also improves with respect to 2NF only.

These are extremely encouraging results and confirm recent investigations of  3NFs~\cite{Jurg.2009,Roth.2012,Hage.2012}. It must be kept in mind that a correct microscopical extension to the Gorkov approach---to include missing 3NFs in both the anomalous and higher order self-energies---is still missing. This requires substantial work to develop and implement  correctly the formalism  and will be addressed in the coming future.


\paragraph{Acknowledgments}
This work was supported by the UK's STFC Grants ST/I003363 and ST/J00005, by the German DFG through grant SFB 634 and Helmholtz Alliance Program, contract HA216/EMMI, and Canada's NSERC Grant No. 401945-2011.
Calculations were performed using HPC resources from GENCI-CCRT (Grant 2012-050707).



\bibliographystyle{aipproc}   


\IfFileExists{\jobname.bbl}{}
 {\typeout{}
  \typeout{******************************************}
  \typeout{** Please run "bibtex \jobname" to optain}
  \typeout{** the bibliography and then re-run LaTeX}
  \typeout{** twice to fix the references!}
  \typeout{******************************************}
  \typeout{}
 }


\end{document}